\documentclass[11pt]{article}
\usepackage{amsmath,amssymb,color,graphics,epsfig,cite}
\usepackage{graphicx,subfigure}

\usepackage{amsfonts}

\newcommand{\hoch}[1]{$\, ^{#1}$}

\newcommand{\be}{\begin{equation}}
	\newcommand{\ee}{\end{equation}}
\newcommand{\bea}{\setlength\arraycolsep{2pt} \begin{eqnarray}}
	\newcommand{\eea}{\end{eqnarray}}
\newcommand{\nn}{\nonumber}

\def\ft#1#2{{\textstyle{\frac{\scriptstyle #1}{\scriptstyle #2} } }}
\def\fft#1#2{{\frac{#1}{#2}}}

\def\0{{\sst{(0)}}}
\def\1{{\sst{(1)}}}
\def\2{{\sst{(2)}}}
\def\3{{\sst{(3)}}}
\def\4{{\sst{(4)}}}
\def\5{{\sst{(5)}}}
\def\6{{\sst{(6)}}}
\def\7{{\sst{(7)}}}
\def\8{{\sst{(8)}}}
\def\sst#1{{\scriptscriptstyle #1}}

\begin{document}
	
\begin{center}

{\Large {\bf First Law of Black Hole Interior Dynamics}}
		
\vspace{20pt}
		
Ze-Xuan Xiong\hoch{1} and H. L\"u\hoch{2,1,3}
		
\vspace{10pt}

{\it \hoch{1}The International Joint Institute of Tianjin University, Fuzhou,\\ Tianjin University, Tianjin 300350, China}

\medskip

{\it \hoch{2}Center for Joint Quantum Studies, Department of Physics,\\
School of Science, Tianjin University, Tianjin 300350, China }

\medskip

{\it \hoch{3} Peng Huanwu Center for Fundamental Theory, Hefei, Anhui 230026, China}

\vspace{40pt}
		
\underline{ABSTRACT}
\end{center}

We obtain a local first law of dynamics within the interior geometries that are governed by cosmological solutions connecting the event horizon and the Kasner-like singularity. Evaluating the Iyer-Wald identity between the horizon and near-singularity geometries, we define a Kasner potential and transported response coefficients. We test the first law by both exact and numerical solutions. Our formalism provides a self-consistent approach to study the interior dynamics without having to appeal to the outside geometry and the asymptotic charges. It may provide a new approach to study the phase structures of the interior dynamics.

\vfill {\footnotesize xzx\_167@tju.edu.cn\ \ \ mrhonglu@gmail.com}
	
	\thispagestyle{empty}
	\pagebreak
	
	




\section{Introduction}\label{sec1}

The first law of black hole thermodynamics \cite{Bardeen:1973gs,Hawking:1974rv} relates variations of global charges to the response of the event horizon. This exact 1-form differential structure is usually formulated outside the black hole, where temperature, entropy, mass and conserved charges are defined either on the horizon or at the asymptotic infinity. The interior cosmological region, however, contains an independent dynamical system  \cite{Regge:1957td,Hawking:1970zqf, Ruffini:1971bza,Belinsky:1970ew, Belinskii:1973sud,Henneaux:2007ej}, and whether this interior mechanics admits a thermodynamics-like exact 1-form variational principle remains an open question and has been rarely addressed.

A natural arena for studying this issue is Einstein gravity coupled to a scalar field. In spherical symmetry, the scalar hair drives the cosmological interior geometry toward a Kasner-like singularity\cite{Kasner:1921zz}, where the metric and scalar field are characterized by a small set of exponents and amplitudes, see e.g.~\cite{Hartnoll:2020fhc, Liu:2021hap,Henneaux:2022ijt,An:2022lvo,Auzzi:2022bfd,Liu:2022rsy,Gao:2023zbd,Cai:2023igv,
Albrychiewicz:2024gti,Zhang:2025tsa,Zhang:2025hkb,Xiong:2026npi,Chew:2026mus,
Zhao:2026mkx,Gao:2026tck}. These data encode the final stage of classical interior evolution, but their relation to horizon quantities and scalar charges has not been precisely formulated. A recent attempt of finding algebraic relations between the data from the Kasner region and black hole asymptotic data such as the mass and scalar charge was initiated \cite{Xiong:2026npi}.

In this paper, we establish a precise first law for the black hole interior dynamics. We define an interior ``Kasner potential'' $M_{\rm K}$ from the Kasner region and show that its variation over a family of electrically-charged scalar-hairy black holes takes the form
\be\label{firstlaw}
\delta M_{\rm K} = {\cal T} \delta S  +  {\it \Phi}_e \delta Q_e + {\phi}_{\rm K} \delta \Sigma_{\rm K}\,,
\ee
where the notations will be apparent in our explicit derivation, which uses the covariant phase-space/Iyer-Wald formalism \cite{Wald:1993nt,Iyer:1994ys}, evaluated on a hypersurface extending from the horizon to the Kasner region. The resulting identity is then tested by both analytic and numerical solutions.

This first law suggests that black hole singularities carry thermodynamic response data constrained by the exterior solution. It provides a bridge between horizon data and Kasner dynamics and gives a concrete framework for studying how the interior structure varies under changes of entropy and the scalar charge. The paper is organized as follows. In section 2, we study the interior parameters from both the near-horizon and Kasner geometries. In section 3, we apply the Iyer-Wald formalism and derive the first law \eqref{firstlaw}. In section 4, we test the first law with various solutions. In section 5, we address the issue of Smarr relations. We conclude the paper in section 6.

\section{Parameters of black hole interior}

To develop a general framework for the interior dynamics of charged, scalar-hairy configurations, we work with a general class of Einstein-Maxwell-Scalar (EMS) theories that allow for a generic potential $V(\phi)$, described by the action
\be
\label{eq:EMD-action}
I = \frac{1}{16\pi}\int d^{4}x\sqrt{-g}\Bigl[R- \frac{1}{2} (\nabla\phi)^{2}
- \frac{1}{4}Z(\phi) F^2- V(\phi)\Bigr] \,,\qquad F=dA\,,
\ee
where there is a non-minimal coupling $Z(\phi)$ between the scalar and the Maxwell field. We consider the ansatz to access both the horizon and the singularity, which takes the form
\be
\label{eq:EMD-ansatz}
ds^2=-h(r)dt^2+\frac{dr^2}{f(r)}+r^2d\Omega_2^2\,,\qquad \phi=\phi(r),\qquad A=A_t(r)dt\,.
\ee
Assuming horizon is at $r=r_h$, for the outside $r>r_h$, it is the most general spherically-symmetric and static ansatz in the areal-radius gauge. For the interior $r<r_h$, the solution is cosmological with $(r,t)$ becoming time and space respectively. The temperature, entropy, and the electric potential on the horizon are given by
\be
T=\frac{\sqrt{h^{\prime}(r_h)f^{\prime}(r_h)}}{4\pi}\,,
\qquad
S=\pi r_h^2\,, \qquad \Phi_{\rm H}=A_t(r_h)\,.
\ee
From the action \eqref{eq:EMD-action} and the static ansatz \eqref{eq:EMD-ansatz}, we define the local charges enclosed by a 2-sphere \(S_r^2\) of radius \(r\) as
\be
Q_e \equiv \frac{1}{16\pi} \int_{S_r^2} Z(\phi)\,{*F} \,, \qquad
Q_\phi(r) \equiv -\frac{1}{16\pi} \int_{S_r^2} \iota_{\partial_t}(*d\phi) \,,
\ee
where \(\iota_{\partial_t}\) denotes contraction with Killing vector $\partial_t$.
The Maxwell charge \(Q_e\) is radially conserved by the Maxwell equation \(d[Z(\phi){*F}]=0\), while the scalar momentum \(Q_\phi(r)\) is generally not conserved and serves as a local canonical momentum conjugate to \(\phi(r)\) at radius \(r\).

With the horizon dynamic quantities in hand, we now extract the dynamic quantities at the singularity. For the scalar-hairy configurations considered here, no inner horizon forms \cite{Xiong:2026npi}, and the interior approaches a Kasner-type spacelike singularity. Near the singularity, the geometry takes the form
\be
ds^2 = -d\tau^2 + a_1^2 \tau^{2 P_t}dt^2 + a_2^2 \tau^{2P_T}d\Omega_2^2\,, \quad \, \phi=2 P_\phi \log{\tau}\,,
\ee
where $(P_t,P_T,P_\phi)$ are Kasner exponents that meet the two Kasner constraints $P_t+2P_T=1$ and $P_\phi^2+3P_T^2-2P_T=0$, provided that the scalar kinetic term dominates in the Kasner region. Recasting this metric into the areal gauge \eqref{eq:EMD-ansatz} identifies $r\sim\tau^{P_T}$. Consequently, in the Kasner region, the metric functions and the scalar behave as
\be
\label{eq:near-singularity-expansion}
f(r)\sim -F_0\,r^{-l}\,,\qquad h(r)\sim-H_0\,r^{l-2}\,,\qquad \phi(r)\sim c_1\log r+\phi_{\rm K} \,,
\ee
with $l\equiv2/P_T - 2$, a consequence of the first Kasner constraint. The second Kasner constraint further requires  $l=1+c_1^2/4$, relating the metric power-law behavior at the Kasner singularity to the scalar hair. This metric power-law behavior allows us to define a set of parameters characterizing the singularity, including the Kasner exponents $(P_t,P_T,P_\phi)$ and the coefficients $(F_0 ,H_0)$ that describe the leading behavior of the metric functions. It is useful to define the following combination in the Kasner region \cite{Xiong:2026npi}
\begin{align}
\Xi=-\lim_{r\rightarrow0^+}r f(r)\sqrt{\frac{h(r)}{f(r)}}=\sqrt{F_0\,H_0}\,.
\end{align}
This limit is well defined for the scalar kinetic-dominated Kasner configurations considered here, as guaranteed by the first Kasner constraint.

The local quantities defined above acquire the values at the singularity by taking the Kasner limit \(r\to0^+\). The electric potential and the local scalar charge both have well-defined values in the Kasner singularity limit, given by
\be
\Phi_{\rm K} \equiv \lim_{r\to0^+}A_t(r)=A_t(0) \,,\qquad
Q_\phi^{\rm K} \equiv\lim_{r\to0^+}Q_\phi(r) =
\frac14\Xi c_1\,,
\ee
For later convenience, we define the normalized scalar charge in the Kasner region
\begin{align}
\Sigma_{\rm K} \equiv \frac{Q_\phi^K}{\Xi} = \frac{c_1}{4}.
\end{align}
Then, combining this normalized scalar charge with the near-singularity geometric parameter $F_0$ , we define the Kasner potential as
\be
\label{Kasner Mass}
M_{\rm K} = \lim_{r\rightarrow 0^+} \Big(\log\sqrt{F_0} + \Sigma_{\rm K}\,(\phi(r)-c_1\log r)\Big)=\ft12 \log F_0 + \phi_{\rm K} \Sigma_K\,.
\ee
(Note that in natural unit $G=c=\hbar =1$, $F_0$ is dimensionless, as the Plank length $\ell_p=1$.) The quantity $M_{\rm K}$ should not be interpreted as a conventional energy or ADM mass. It is instead a dynamic parameter in the Kasner region, defined through the Iyer-Wald formalism, which we shall discuss in the next section.

\section{Iyer-Wald derivation of the interior first law}

We now derive a precise first-order differential relation among the data from the horizon and the Kasner region using the Iyer-Wald formalism. For a general variation of the fields \(\Psi=\{g_{\mu\nu},A_\mu,\phi\}\), the Lagrangian four-form satisfies \cite{Wald:1993nt,Iyer:1994ys}
\be
\delta \boldsymbol L = \boldsymbol E_\Psi \delta\Psi +
d\boldsymbol\Theta(\Psi,\delta\Psi)\,,
\ee
where $E_\psi=0$ gives the equations of motion. For a Killing vector \(\xi\), we define the 2-form \cite{Wald:1993nt,Iyer:1994ys}
\begin{align}
\boldsymbol k_\xi(\delta\Psi;\Psi)
=\delta\boldsymbol Q_\xi-\iota_\xi\boldsymbol\Theta(\Psi,\delta\Psi)\,,
\end{align}
where \(\boldsymbol Q_\xi\) is the Noether charge 2-form, and the variation applies only on the integration constants of a solution. By virtue of Stokes' theorem, the Wald formalism implies that if there is no singularity between the horizon $r_h$ and the regulated Kasner surface $r=\epsilon\rightarrow 0$, there exists an identity $\delta {\cal H}_\xi(r_h)=\delta {\cal H}_\xi(\epsilon)$, where
\bea
\delta {\cal H}_\xi(r)
&\equiv& \int_{S_r^2}\boldsymbol k_\xi =\frac{1}{16\pi}\int_{S_r^2} \bigg\{r\Big( -2\sqrt{\frac{h(r)}{f(r)}}\,\delta f(r) - f(r)\, \sqrt{\frac{h(r)}{f(r)}}\,r\,\phi^{\prime} \delta \phi \Big)\nn\\
&&\qquad\qquad- A_t(r)\, \delta \Bigl(r^2 \sqrt{\frac{f(r)}{h(r)}}\, Z(\phi) \,A_t^{\prime}(r)\Bigr)\bigg\}\,.
\eea
This is mathematically true despite the fact that the interior is cosmological and $r$ is a time coordinate. Evaluating $\delta {\cal H}$ on the horizon, we have
\be
\delta {\cal H}_\xi(r_h) = T\,\delta S+\Phi_{\rm H}\,\delta Q_{e} \,.
\ee
At the regulated Kasner surface $r=\epsilon$, $\delta {\cal H}$ has a potential logarithmic divergence, given by $\delta H_\xi(\epsilon)=\Xi (c_1 \delta c_1 - 2 \delta l) \log(\epsilon) + \cdots$. This could-be divergent term vanishes identically owing to the Kasner constraint given under \eqref{eq:near-singularity-expansion}. The finite ellipses are
\be
\delta H_\xi(\epsilon) = \Xi\,\delta\log\sqrt{F_0} +
\Phi_{\rm K} \,\delta Q_{e} + \Xi \Sigma_{\rm K}\,\delta \phi_{\rm K} + {\cal O}(\epsilon)\,,
\ee
Equating the $\delta {\cal H}$'s on the two surfaces, we have
\be\label{IW:identity}
\delta\log\sqrt{F_0} + \Sigma_{\rm K}\,\delta \phi_{\rm K} = \frac{T}{\Xi}\delta S +
\frac{\Phi_H-\Phi_K}{\Xi}\delta Q_{e} \,.
\ee
We therefore define the following two scaled parameters,
\be
{\cal T}\equiv\frac{T}{\Xi}\,,\qquad
{\it \Phi}_e\equiv\frac{\Phi_H-\Phi_K}{\Xi}\,,
\ee
Using the Kasner potential \eqref{Kasner Mass}, the differential relation \eqref{IW:identity} leads to the first law \eqref{firstlaw}. It is important to note that in the static ansatz \eqref{eq:EMD-ansatz}, the constant time rescaling $t\rightarrow \lambda t$ renders $(h, A_t)$ functions ambiguous. For asymptotically-flat geometries, this ambiguity is fixed by requiring a specific choice of the speed of light at infinity; however, such fiducial reference is lacking if we focus on the interior region only. Intriguingly, our definitions of $M_{\rm K}$, ${\cal T}$ and ${\it \Phi}_e$, and hence the first law are independent of the $t$ rescaling parameter $\lambda$.

Notably, our first law \eqref{firstlaw} is a local relation in the cosmological region between the horizon and the Kasner singularity. Its derivation uses only the $r$-conservation of the Iyer-Wald Hamiltonian variation, without inferring any property outside the horizon. The first law is therefore applicable for more general black interior, defined as cosmological region surrounded by a horizon, instead of being restricted to the black hole interior only. Specifically, for black holes, defined as having both horizon and proper asymptotic Minkowski spacetime, the scalar hair parameter is not an independent parameter, but a function of $(S,Q_e)$ in the canonical ensemble. This would imply that even though the first law holds, the three parameters $(S, Q_e, \Sigma_K)$ describing the black hole interior are not independent. For the more general black interior, all the three parameters are independent and hence the first law \eqref{firstlaw} becomes more nontrivial.

\section{Analytic and numerical tests of the interior first law}

In this section, we use explicit solutions to verify the first law \eqref{firstlaw} of black interior derived in the previous section. We consider both exact and numerical solutions. For a given interior solution, the relevant quantities we need to calculate involves $\{T,S\}$, which can be evaluated on the horizon solely. The quantities $\{\phi_{\rm K}, \Sigma_{\rm K}, {\it \Phi}_e, M_{\rm K}, \Xi\}$ should be evaluated in the Kasner region. The conserved electric charge $Q_e$ is given independent of $r$.

\subsection{Testing the first law with exact black hole solutions}

Exact solutions provide analytic calibrations of the relevant data in the first law from the Kasner regulated surfaces. The simplest example is perhaps the Schwarzschild black hole of mass $m$. We find that $M_{\rm K}=\fft12 \log(2m)$, $T=1/(8\pi m)$ and $S=4\pi m^2$. Since $\Xi=2m$, it is easy to verify the interior first law $\delta M_{\rm K} = {\cal T} \delta S$ is satisfied.

\paragraph{Charged EMD black holes.} We first specialize the theory to the standard Einstein-Maxwell-Dilaton (EMD) model, with $V(\phi)=0$ and $Z(\phi)=e^{a\phi}$, where $a$ is the dilaton coupling constant. Charged black holes in EMD theory were constructed in \cite{Gibbons:1987ps,Garfinkle:1990qj}. Defining $n=4/(1+a^2)$, the solution is
\be
ds^2 = -H^{-n/2}f_0\,dt^2 + H^{n/2} \left(\frac{dx^2}{f_0}+x^2d\Omega_2^2\right)\,,\quad
\phi=\frac{na}{2} \log H\,,\qquad A_t=\fft{\sqrt{nq(m+q)}}{x+q}\,,
\ee
where $f_0=1-m/x$ and $H=1+q/x$. The black hole horizon is located at $x=m$ and the spacetime singularity is at $x=0$. To study the cosmological interior in $x\in (0,m)$, it is advantageous to convert the solution to the areal-gauge radius $r= xH^{n/4}$.
A simple but careful evaluation gives us
\bea
&&M_{\rm K} = \frac{\log m}{2} - \frac{\log q}{2a^2}\,,\qquad
Q_e = \ft14 \sqrt{n q (m+q)}\,,\qquad {\it \Phi}_e = -\fft{1}{a^2 Q_e}\,,
\qquad \Sigma_{\rm K}=-\frac{1}{2a}\,,\nn\\
&& {\cal T} =\fft{m}{4\Xi\,S} \,,\qquad S=\pi m^2 (1+q/m)^{\fft12n}\,,\qquad
\Xi=\ft14 m (4-n)\,,\qquad \phi_{\rm K} = \fft2{a}\log q\,.
\eea
It is now straightforward to verify that the first law \eqref{firstlaw} is satisfied.

\paragraph{Neutral scalar hairy black holes.} Turning off the Maxwell field, exact scalar hairy black holes can be constructed for a suitable class of scalar potentials \cite{Anabalon:2012ta,Anabalon:2013qua,Gonzalez:2013aca,Acena:2013jya,Feng:2013tza}. The interior structure of these black holes was studied in \cite{Xiong:2026npi}. Here, we derive their first law of the interior dynamics. The black hole solutions take the form
\cite{Feng:2013tza}
\bea
ds^2 &=& -\fft{B(x)}{H(x)^{\mu}}\,dt^2 + \frac{H(x)^{\mu} dx^2}{B(x)} +
    x^2 H(x)^{1+\mu} d\Omega_2^2\,,\qquad \phi(x) =\sqrt{1-\mu^2} \log H(x)\,,\nn\\
B(x) &=& 1+ \frac{\alpha x^2}{2\mu(1-4\mu^2)}\Big(H^{2\mu+1}-\mu(2\mu+1)H^{2\mu}
+(4\mu^2-1)H-\mu(2\mu-1)\Big)\,,
\eea
where $H=1+q/x$, and the parameter $\mu \in [-1,1/2)$ and $\alpha$ are the coupling constants of the theory. The solutions have only one integration constant $q$, and they describe black holes provided that
\be
\Delta \equiv \frac{\alpha q^2}{2(1-2\mu)}-1>0\,.
\ee
The horizon is located at the only positive root of $B(x)$, where we can determine the temperature $T$ and entropy $S$  using the standard method. We find that
\be
T\delta S= \frac{2\mu+\alpha q^2}{4}\,\delta q\,,
\ee
As was shown in \cite{Xiong:2026npi}, the Kasner singularity is located at $x=0$. It is now straightforward to obtain the relevant data from the regulated singularity surface. We find that
\bea
&&F_0=\frac{(1-\mu)^2}{4}\,\Delta\,q^{2/(1-\mu)}\,,\qquad H_0 =\Delta\,q^{-2\mu/(1-\mu)}\,,
\qquad \Xi=\frac{1-\mu}{2}\,\Delta q\,,\nn\\
&&M_{\rm K}=\log\sqrt{F_0} + \phi_{\rm K} \Sigma_{\rm K}\,,\qquad  \Sigma_{\rm K} = -\frac12 \sqrt{\fft{1+\mu}{1-\mu}}\,,\qquad \phi_K =  2 \sqrt{\fft{1+\mu}{1-\mu}}\log q\,.
\eea
It can be easily verified that the first law \eqref{firstlaw} is satisfied.

It is worth commenting that both above examples are black holes and the no-scalar-hair theorem implies that the scalar charge $\Sigma_{\rm K}$ extracted from the Kasner singularity is not an independent parameter. It turns out that it depends on the coupling constants only, in other words, we have $\delta \Sigma_K=0$, rendering its contribution to the first law as being trivial. In order to see the full effect of the first law, we need to relax the black hole condition, by including solutions with a horizon that may not be integrated smoothly to asymptotic Minkowski infinity. We are unaware of such exact solutions and we shall construct numerical examples next.

\subsection{Testing the first law with numerical horizon-to-Kasner geometries}

For simplicity, we shall turn off the Maxwell field and consider only the Einstein-scalar theory. We consider two different scalar potentials, $V_1=g_5 \phi^5 + g_7 \phi^7$ and $V_2=\ft12 g_2 \phi^2$. There can be no scalar-hairy black holes for $V_2$, but they exist for $V_1$, in which case, a large number of scalar hairy black holes were constructed in our earlier paper \cite{Xiong:2026npi}. However, horizons exist for both potentials and the general near-horizon geometry in the areal gauge is given by
\bea
f(r) &=& f_1 (r-r_0) +  f_2 (r-r_0)^2 + \cdots\,, \qquad
h(r)=  (r-r_0) +  h_2 (r-r_0)^2 + \cdots\,, \nn \\
\phi(r) &=& \phi_0 +  \phi_1 (r-r_0) +  \phi_2 (r-r_0)^2 + \cdots\,.
\eea
The coefficients of the Taylor series expansions can be solved order by order in terms of the powers of $(r-r_0)$. The detailed expressions of several leading order results were given in \cite{Xiong:2026npi}. We can in principle obtain all the coefficients in terms of two parameters, the horizon radius $r_0$ and scalar hair $\phi_0$ on the horizon. For numerical solutions, we choose $g_2=g_5=g_7=1$ and let $r_0\in(1,4)$ and $\phi_0\in(0.1,0.5)$. It is worth commenting again that none of these horizon geometries can be smoothly integrated to asymptotic Minkowski infinity for $V_2$. For $V_1$, $\phi_0$ has to be fine-tuned to be a specific function of $r_0$ to become an asymptotically-flat black hole.

\begin{figure}
    \centering
    \includegraphics[width=0.45\linewidth]{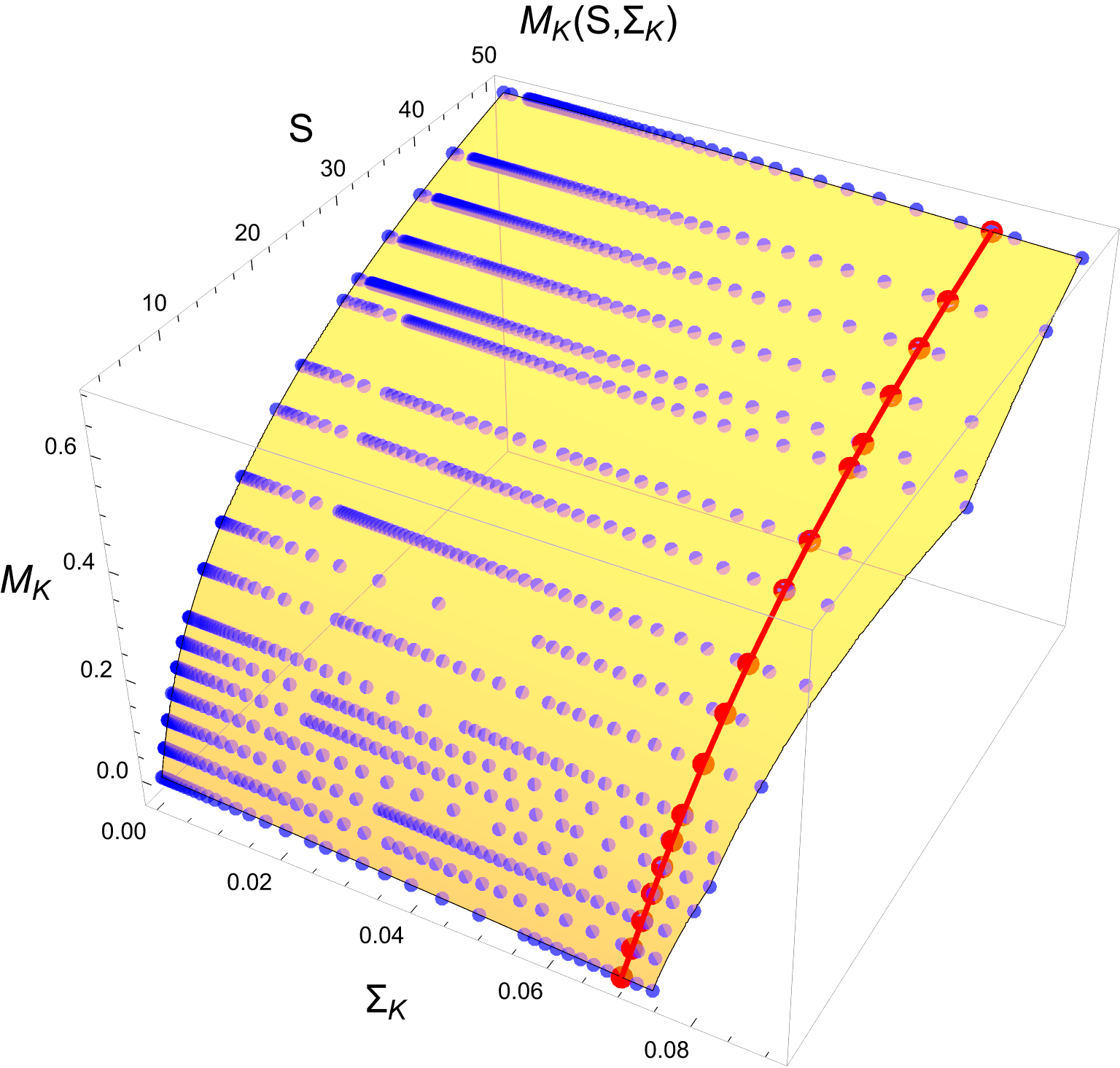}\label{fig:MSQg5}\ \ \
    \includegraphics[width=0.45\linewidth]{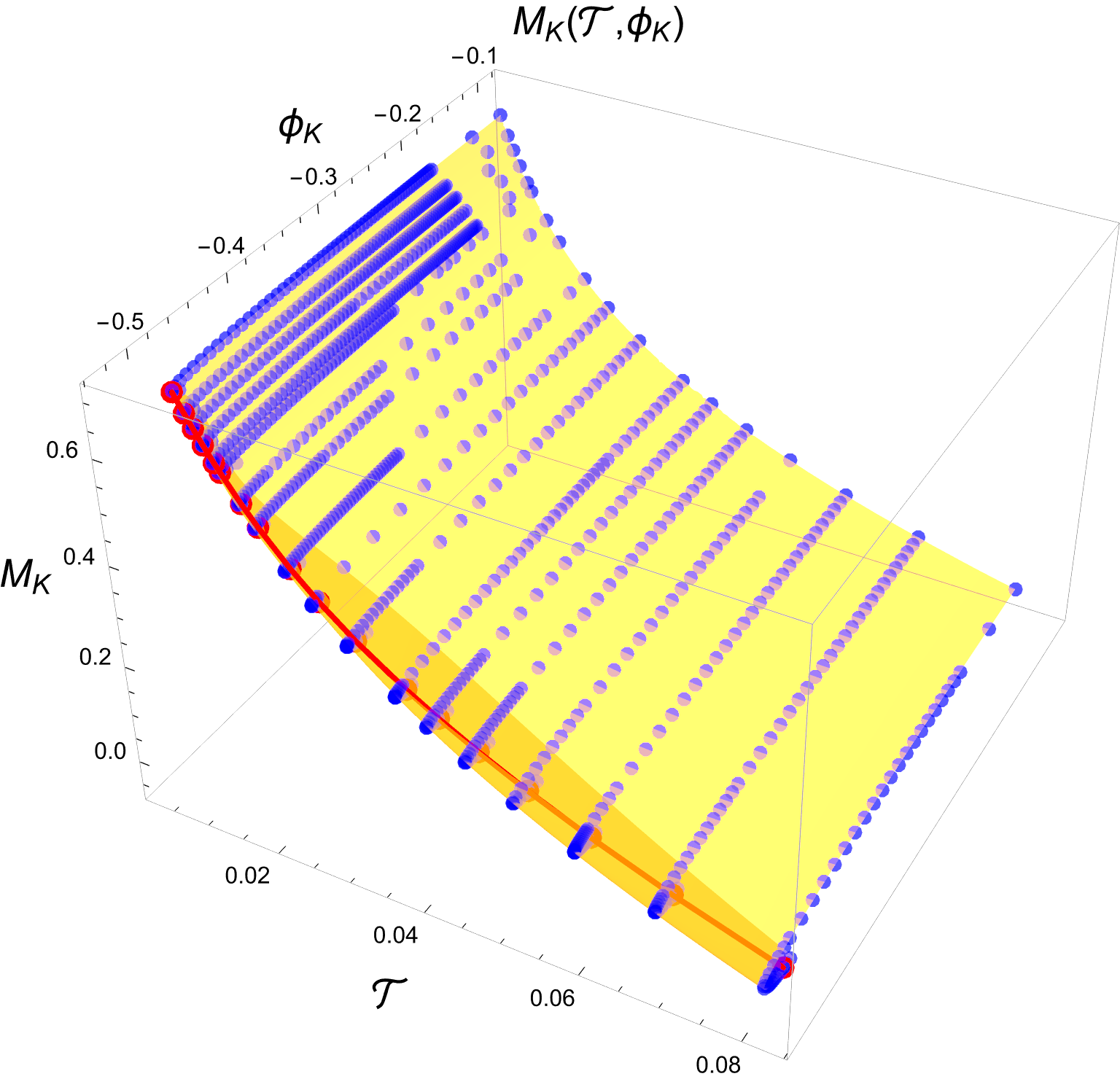}\label{fig:MTPg5}\ \ \
    \includegraphics[width=0.45\linewidth]{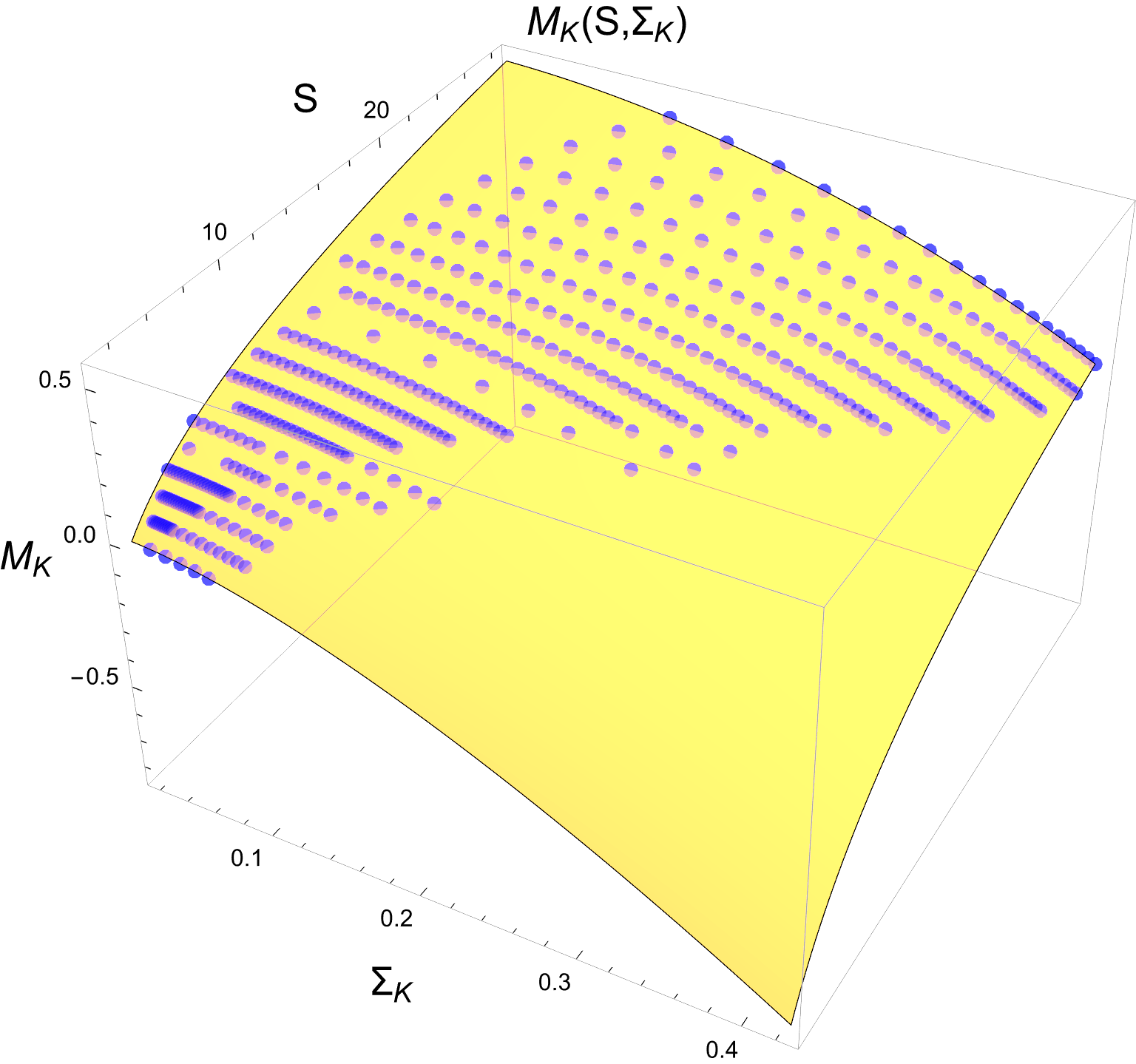}\label{fig:MSQg2}\ \ \
    \includegraphics[width=0.45\linewidth]{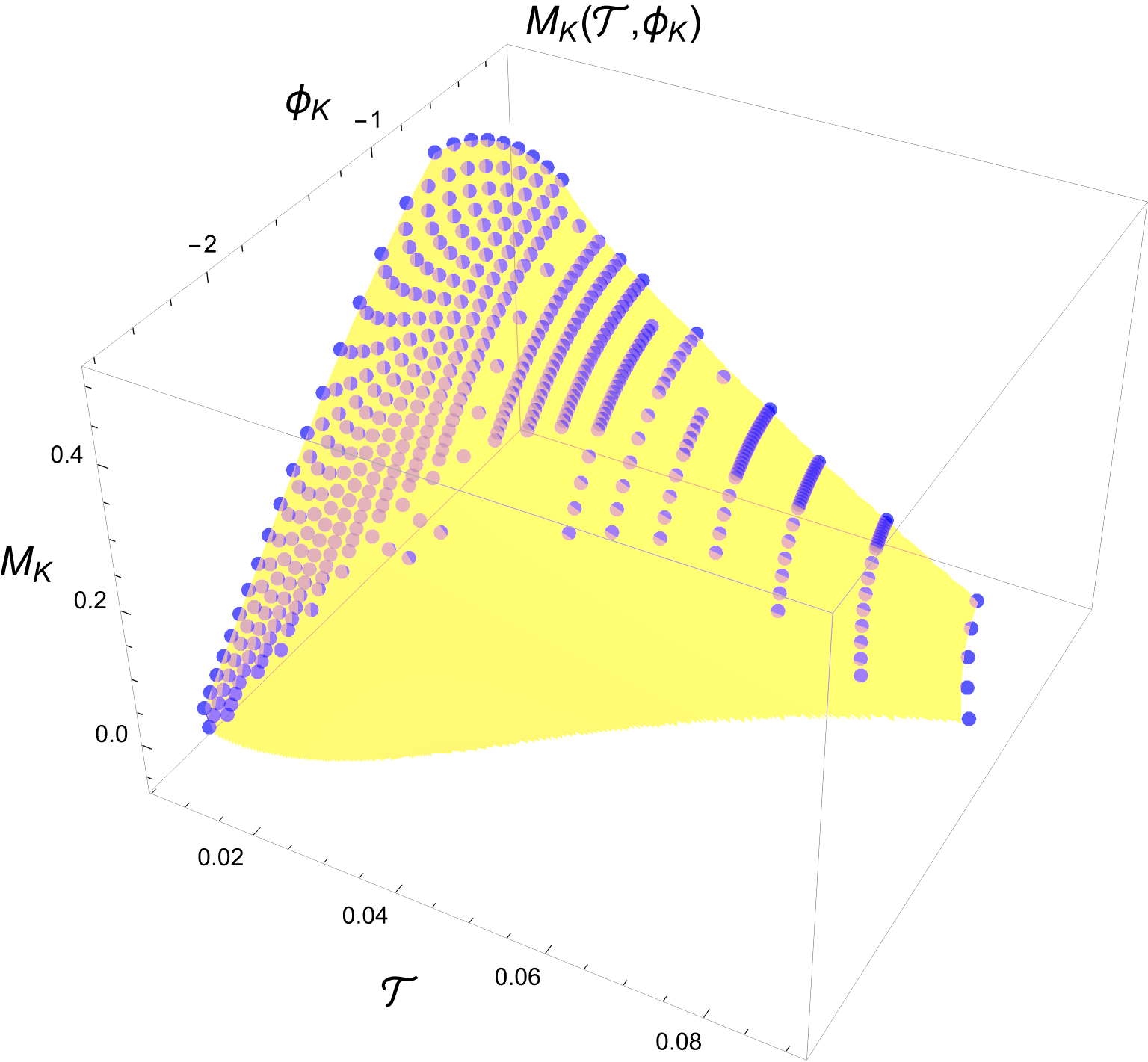}\label{fig:MTPg2}
    \caption{\small Here we present the numerical testing of the first law of the interior dynamics for Einstein-scalar theory with $V_1$ (top two panels) and $V_2$ (bottom two panels) potentials. The yellow surfaces are data fitting curves and the dots are explicit numerical data. The red line and red dots are associated with the data of a black hole interior, which exist only for the $V_1$ potential. As explained in the text, their matching demonstrates that the first law holds for the general two-parameter interior cosmological solutions.}
\label{fig:overall}
\end{figure}

However, these is no obstacle to solving the equations of motion and integrating from the general horizon to the Kasner region at $r\rightarrow 0$. We therefore obtain a general two-parameter family of cosmological solutions of the black interior. We can extract numerically all the dynamic variables associated with the first law. In order to test the first law of the Einstein-scalar theory, namely $\delta M_{\rm K} = {\cal T} \delta S + \phi_{\rm K} \delta \Sigma_{\rm K}$, we first perform data fitting to extract the relation $M_{\rm K}=M_{\rm K}(S, \Sigma_K)$, which allows us to calculate the ``theoretical'' results based on the first law
\be
{\cal T}^{\rm th} = \fft{\partial M_{\rm K}}{\partial S}\,,\qquad \phi_{\rm K}^{\rm th}=
\fft{\partial M_{\rm K}}{\partial \Sigma_{\rm K}}\,.
\ee
The test of the first law therefore becomes comparing the theoretical $M_{\rm K}({\cal T}^{\rm th}, \phi_{\rm K}^{\rm th})$ to the numerical data $M_{\rm K}({\cal T}, \phi_{\rm K})$. We find they fit perfectly, as shown in Fig.~\ref{fig:overall}.  In particular, the red line and red dots are associated with the interior of the black hole solutions, which exist for $V_1$, but not $V_2$. However, the general interior solutions may not be related to a black hole, but the first law of black interior holds nevertheless.

\section{The subtlety of the Smarr relation}

Our first law \eqref{firstlaw} has a virtue that its formalism is independent of a constant rescaling of the coordinate $t$, which is important as we are lacking of a fiducial frame for either the horizon or Kasner singularity. It has a drawback that the {\it usual} Smarr relation typically accompanied to the first law of black hole thermodynamics no longer holds. We address this here using both Euler derivation and the generalized Komar formalism.

\subsection{Euler derivation and logarithmic scaling}

When the dimensionful potential $V(\phi)$ is absent, our theory \eqref{eq:EMD-action} and the dynamic quantities involved in the first law \eqref{firstlaw} obey the following scaling
\be
S\rightarrow \lambda^2 S\,,\qquad Q_e\rightarrow \lambda Q_e\,,\qquad \Sigma_{\rm K}\rightarrow \Sigma_{\rm K}\,,\qquad F_0\rightarrow \lambda^l F_0\,,\qquad
\phi_{\rm K} \rightarrow \phi_{\rm K} - c_1 \log\lambda\,.
\ee
It follows that the Kasner potential $M_{\rm K}$ is shifted by a logarithmic term, namely
\be
M_{\rm K} (\lambda^2 S, \lambda Q_e, \Sigma_{\rm K}) = M_{\rm K} (S,Q_e, \Sigma_{\rm K})
 + (\ft12 -2 \Sigma_K^2)\log\lambda\,,
\ee
where we have used the identity $\Sigma_{\rm K} = c_1/4$ and $l= 1 +4 \Sigma_{\rm K}^2$. Thus the Kasner potential is very different from the ADM mass that leads to the standard Smarr relation. Differentiating with $\lambda$ at $\lambda=1$ gives the Euler operator identity
\be
\Big(2S\fft{\partial}{\partial S} + Q_e \fft{\partial}{\partial Q_e}\Big) M_{\rm K} =\fft12 -2 \Sigma_K^2\,.
\ee
Using the first law \eqref{firstlaw}, we arrive at the new Smarr-like relation
\be
2{\cal T} S + {\it \Phi}_e Q_e = \fft12 -2 \Sigma_K^2\,.\label{smarr1}
\ee
For the Schwarzschild black hole, we have $\Sigma_K=0$, $\Xi=2m$, and this relation becomes $2T S=\Xi/2=m$, the standard Smarr relation. For the EMD black holes we discussed earlier, it can be easily verified that the above relation leads to the standard Smarr relation.

\subsection{Generalized Komar formalism}

When the potential $V$ is not zero, the situation becomes more complicated, since $V$ is not invariant under the scaling discussed above. For Killing vector $\xi=\partial_t$, its 1-form is $\boldsymbol\xi=-h(r)\,dt$. The Komar charge of a radius-$r$ sphere and its derivative are
\be
K(r)\equiv-\frac{1}{8\pi}\int_{S_r^2}*d\boldsymbol\xi
=\frac{r^2}{2}\sqrt{\frac{f(r)}{h(r)}}\,h'(r)\,,\qquad
K'(r) = -Q_{e}A_t'(r) -\frac12\,r^2\sqrt{\frac{h}{f}}\,V(\phi) .
\ee
We can introduce the generalized Komar charge
\be
J(r) = K(r) + Q_e A_t(r)\,,\qquad\hbox{obeying}\qquad
J'(r)= -\frac12\,r^2\sqrt{\frac{h}{f}}\,V(\phi)\,.
\ee
Its covariant form language is simply $d\boldsymbol{{\cal J}_\xi}=-\fft{1}{8\pi} V(\phi)\, {*\boldsymbol{\xi}}$. When $V=0$, the Smarr relation is simply the consequence of Stokes' theorem applied on radially conserved $J(r)$, which reads
\be
r\rightarrow r_0:\quad J_{\rm H}=2 T S + \Phi_{\rm H} Q_e\,;\qquad
r\rightarrow 0:\quad J_{\rm K} = (1-\ft12l) \Xi + \Phi_{\rm K} Q_e\,.
\ee
Setting $J_{\rm H} = J_{\rm K}$ and dividing the equation by $\Xi$, we arrive at \eqref{smarr1}. When $V$ is non-vanishing, we have a more general Smarr-like relation, given by
\be
2\mathcal{T} S
+\mathit\Phi_e Q_{e}
+\frac{1}{2\Xi}\int_0^{r_h}dr\,r^2\sqrt{\frac{h}{f}}\,V(\phi)
=\frac12-2\Sigma_{\rm K}^2\,.
\ee
It is easy to see that this relation is independent of the constant rescaling of $t$. If the solution turns out to be the interior of a black hole, the above Smarr-like relation reproduces the black hole Smarr relation involving a scalar potential. For example, for the Schwarzschild-AdS black hole of mass $M$, we have $\Sigma_K=0$ and $\Xi=2M$ and $V=2\Lambda$ where $\Lambda$ is the cosmological constant. The above relation becomes the stanard Smarr relation of the Schwarzschild-AdS black hole, namely $M=2 T S + \ft13 \Lambda r_h^3$, where $r_h$ is the horizon radius.

\section{Conclusions}

In this paper, we formulated a local first-order variation principle for dynamics of black interior that is described by cosmological solutions connecting a horizon to Kasner singularity. The geometry outside the horizon is spherically-symmetric and static associated with ansatz \eqref{eq:EMD-ansatz}. If the outside solution also connects the horizon to a smooth asymptotic infinity, it describes a black hole. However, the more general interior cosmological solutions may not all be related to black holes, and we refer them generally as black interiors.

We formulated the first law for a general class of EMS theories \eqref{eq:EMD-action} with a general $Z(\phi)$ and $V(\phi)$. The theory admits many exact black hole solutions that allowed us to test the first law analytically. In order to see the nontrivial effect of the scalar charge to the first law, we also constructed numerical horizon-to-Kasner solutions where the scalar hair could be an independent parameter. We found that both analytic and numerical solutions confirmed the first law.

Our formulation was based on the Iyer-Wald formalism. Although we have well-prescribed independent methods to compute $(S,Q_e,\Sigma_K)$, the Kasner potential $M_{\rm K}$ is solely determined by the Iyer-Wald formalism so that we can have a closed 1-form such that the differential relation can be expressed as a first law, analogous to that of black hole thermodynamics. The Kasner potential is very different from the black hole ADM mass and the Smarr relation must be readdressed. Interestingly, we found that the modified interior Smarr relation was actually equivalent to the usual black hole Smarr relation if the solution actually describes the interior of a black hole. Our first law thus may provide a new venue to characterize the phase structures of black interior dynamics.

It is important to note that when we focus on the interior, the coordinate $t$ is no longer time and there is no obvious normalization for this coordinate. One salient feature of our formulation of the first law is that it is independent of its constant rescaling, making the formalism well-defined without having to appeal to the ``outside'' geometry. It is also worth noting that not all matter fields necessarily admit horizon geometries. For example, Einstein gravity with a free massless scalar can have no horizon at all in spherically-symmetric and static configurations with the scalar hair. We expect that our formalism can also be adopted to derive the first law of dynamics in these spacetimes with unavoidable naked singularities.

\section*{Acknowledgement}

H.L.~is grateful to Peng Huanwu Center for Fundamental Theory for hospitality during the late stage of this work, which is supported in part by the National Natural Science Foundation of China (NSFC) grants No.~12375052, No.~11935009 and No.~12247103, and also by the Tianjin University Self-Innovation Fund Extreme Basic Research Project Grant No.~2025XJ21-0007.

\end{document}